\documentclass[10pt,twocolumn,letterpaper]{article}

\usepackage[pagenumbers]{wacv} % To force page numbers, e.g. for an arXiv version

\usepackage{graphicx}
\usepackage{amsmath}
\usepackage{amssymb}
\usepackage{booktabs}
\usepackage{times}
\usepackage{epsfig}
\usepackage{bm}
\usepackage[section]{placeins}
\usepackage[accsupp]{axessibility}  % Improves PDF readability for those with disabilities.

\newcommand{\tb}[2]{
    \begin{tabular}{c}
        #1\\
        #2
    \end{tabular}
}

\usepackage[pagebackref,breaklinks,colorlinks]{hyperref}

\usepackage[capitalize]{cleveref}
\crefname{section}{Sec.}{Secs.}
\Crefname{section}{Section}{Sections}
\Crefname{table}{Table}{Tables}
\crefname{table}{Tab.}{Tabs.}

\def\wacvPaperID{1299} % *** Enter the WACV Paper ID here
\def\confName{WACV}
\def\confYear{2024}

\begin{document}

\title{ECSIC: Epipolar Cross Attention for Stereo Image Compression}

\author{Matthias Wödlinger\\
TU Wien\\
% {\tt\small mwoedlinger@cvl.tuwien.ac.at}
\and
Jan Kotera\\
UTIA CAS,\\
Czechia
\and
Manuel Keglevic\\
TU Wien
\and
Jan Xu\\
Deep Render
\and
Robert Sablatnig\\
TU Wien
}

% \and
\maketitle

%%%%%%%%% ABSTRACT
\begin{abstract}
   In this paper, we present ECSIC, a novel learned method for stereo image compression. Our proposed method compresses the left and right images in a joint manner by exploiting the mutual information between the images of the stereo image pair using a novel stereo cross attention (SCA) module and two stereo context modules. The SCA module performs cross-attention restricted to the corresponding epipolar lines of the two images and processes them in parallel. The stereo context modules improve the entropy estimation of the second encoded image by using the first image as a context. We conduct an extensive ablation study demonstrating the effectiveness of the proposed modules and a comprehensive quantitative and qualitative comparison with existing methods. ECSIC achieves state-of-the-art performance in stereo image compression on the two popular stereo image datasets Cityscapes and InStereo2k while allowing for fast encoding and decoding.  
\end{abstract}

%%%%%%%%% BODY TEXT
\section{Introduction}
\begin{figure}[t]
	\begin{center}
			\includegraphics[width=0.5\textwidth]{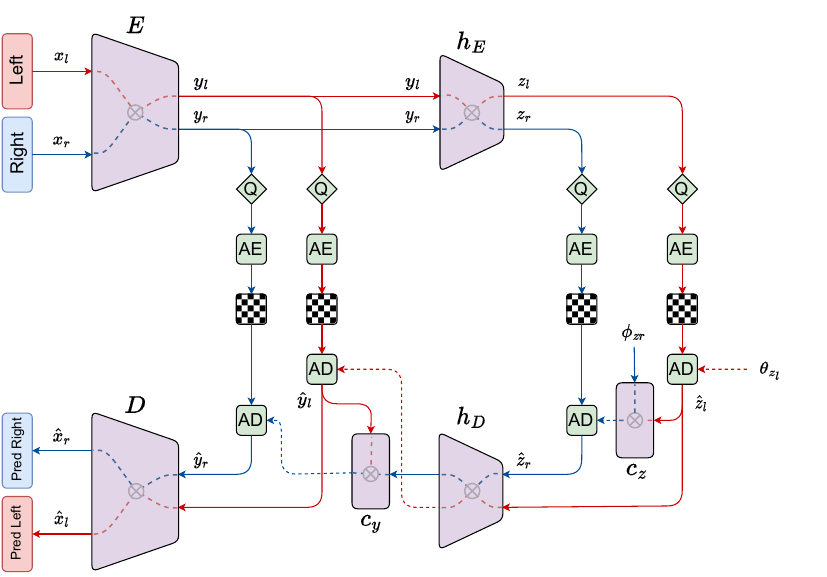}
	\end{center}
	\caption{\label{fig: overview}%
	An overview of the architecture of our ECSIC model. The left and right streams are colored red and blue respectively. The encoder $E$, decoder $D$, hyperprior encoder $h_E$, and decoder $h_D$ are jointly processing the left and right image stream and run in parallel. The stereo context modules $c_y$ and $c_z$ are included only in the right stream and use input from the left stream as contextual information. The Stereo Cross Attention (SCA) modules connect the left and right image stream (depicted as $\otimes$). Submodules in green (quantizers (Q) and arithmetic encoder/decoder (AE/AD)) do not contain any trainable parameters. The bitstreams are denoted with a checkerboard pattern. Dashed lines connecting to AD indicate predicted entropy parameters.}
\end{figure}

Stereo image compression (SIC) aims to jointly compress the two images of the stereo pair more efficiently by exploiting their mutual information and with the same goals as in lossy compression of single images~--~to reduce the bit rate required for storage and transmission while preserving the content and perceived quality. Stereo cameras are used in applications that demand high compression rates and low encoding and decoding latency to facilitate continuous recording or streaming (e.g., in autonomous driving or virtual reality streaming). As such, learned compression methods, which are typically symmetric in encoding and decoding time, offer an advantage over conventional methods, in which encoding is considerably slower. Learned methods transform the image to a latent representation which is then subsequently quantized and entropy coded to the bitstream using a learned probability distribution. They are trained end-to-end by optimizing the weighted trade-off between bitrate and distortion. Optimal rate is achieved by minimizing the cross-entropy of the true latent distribution and its model.
% 1) the optimal warping between left and right images can be highly non-linear

State-of-the-art learned methods already outperform traditional methods, such as BPG, in single image compression~\cite{minnen2018joint}. Images in a stereo pair have high mutual information, optimal compression method should therefore achieve a rate close to that required to compress one of them and thus substantially outperform independent compression of the two images separately. However, occlusions and non-overlapping fields of view between the two cameras in a stereo setup make such a significant reduction of bitrate hard to achieve. To address these difficulties, early learned compression approaches focused on modelling the disparity between the two images to obtain a dense warp field~\cite{Liu2019DSICDS} or a rigid homography transform~\cite{deng2021hesic} to effectively register the two images in order to exploit their similarity. However, such an approach is quite computationally expensive, not to mention that dense disparity map cannot be efficiently transmitted via the bitstream. Later, more efficient approaches appeared that are able to achieve similar performance, such as SASIC~\cite{wodlinger2022sasic}, which combines a simple disparity model (global horizontal shift) with stereo attention to account for smaller local shifts.

In this work, we propose a novel neural network that compresses stereo images without explicit estimation of the disparity warping. The network follows an autoencoder structure with a hyperprior~\cite{balle2018variational} entropy model. The encoder and decoder modules contain a novel stereo attention module that enables joint processing of both images in a stereo pair. Additionally, we introduce two stereo context modules in the entropy model for improved estimation by using the left image as context for the right image. The main contributions of this work are as follows:
\begin{itemize}
    \item We propose ECSIC ({\bf{E}}pipolar {\bf{C}}ross attention for {\bf{S}}tereo {\bf{I}}mage {\bf{C}}ompression), a novel stereo compression method that achieves state-of-the-art performance among SIC models while being fast during both encoding and decoding.
    \item We propose a stereo cross attention module and two stereo context modules for exploiting the mutual information in stereo images for compression and perform an ablation study demonstrating their impact on the overall stereo compression performance.
    \item We evaluate our method quantitatively and qualitatively on two popular stereo image datasets: Cityscapes and InStereo2k.
    \item The method is end-to-end trainable on any stereo image dataset and the code is publicly available \url{https://github.com/mwoedlinger/ecsic}.
\end{itemize}
%%%%%%%%%%%%%%%%%%%%%%%%%%%
%%%%%%%% RELATED WORK %%%%%%%%
%%%%%%%%%%%%%%%%%%%%%%%%%%%%%%
\section{Related Work} 
Based on their principle, image compression methods can be categorized into traditional and learned. In the former, the transformation from the input image to its latent representation is designed by hand; in the latter, it is learned from data by optimizing the rate-distortion loss. Common to both approaches is that the rate savings are ultimately achieved by using an (off-the-shelf) entropy coder that transforms the discrete latent representation to and from the minimum length bitstream. The approximate inverse of the initial transform is then used to reconstruct the decoded image.

\paragraph{Traditional Methods}
The best-known and most widely used image codec is arguably the JPEG method \cite{jpeg} from the 90s. It uses fixed 8x8 block tiling, chroma subsampling, discrete cosine transform, and several next-block prediction modes. Its successor JPEG2000 \cite{jpeg2k} is based on multi-resolution processing with a discrete wavelet transform. The development of modern compression methods has focused on video, and image codecs usually appear as wrappers around intra-frame compression in video codecs such as BPG \cite{bpg} (based on HEVC \cite{hevc}), AVIF \cite{avif} (based on AV1), or VVC-intra \cite{vvc}. The latter has arguably the best compression performance among the traditional methods but is far from being adopted in practice due to its low speed, the lack of production-ready decoders, and restrictive licensing.

\paragraph{Learned Methods}
Pioneering work in learned image compression was done by Toderici et al. \cite{Toderici2016VariableRI} who proposed a recurrent neural network for variable rate image compression. Foundations of the modern approach were laid by Ballé et al. \cite{balle2017end} where an autoencoder-based model with a fixed parametrized distribution of the latent is trained with a rate-distortion loss for a fixed target bitrate. In their subsequent works, the fixed latent entropy model was replaced by a per-pixel Gaussian distribution with parameters predicted for each input image by a separate hyperprior module \cite{balle2018variational} or even an autoregressive context module \cite{minnen2018joint, Mentzer2018ConditionalPM}, which significantly improved performance.

This basic structure was later improved in many ways, for example, by changes in model architecture \cite{cheng2020gm,gao2021iccv,xie2021,he2022}, quantisation approximation \cite{theis2017lossy,guo2021soft}, or theoretical insights into the optimisation problem \cite{yang2020nips}. Much effort has been devoted to improving the context model \cite{minnen2020channel,qian2020,he2021checkerboard, guo2021context,he2022,koyuncu2022}. Recently proposed methods report performance improvements by replacing convolution blocks with transformers or other types of attention modules, mainly in the hyperprior and context model \cite{koyuncu2022,qian2022,kim2022}, but also in the main autoencoder \cite{zhu2022,zou2022}.

A slightly different line of research aims at the realism of image reconstruction in addition to the objective of rate-distortion optimisation. This is most commonly achieved by GANs \cite{tschannen2018gan,agustsson2019generative,mentzer2020high,Gao2021c,he2022po} but also by other generative models such as denoising diffusion \cite{theis2022,Ghouse2023}.

\paragraph{Stereo Image Compression}
In compression of stereo images, bitrate is saved by exploiting the mutual information between the left and right images of the stereo pair. Although somewhat similar to compression of consecutive frames of a video sequence, disparity in stereo images is not well modelled by optical flow, and direct application of video codecs is therefore suboptimal. Of the traditional methods, MV-HEVC \cite{mvhevc} is an extension of the HEVC video codec for multi-view sequences with good performance but lacking support for high bit depth processing and 444 chroma mode. Learnable lossless stereo compression has been proposed by Huang et al. \cite{huang2021l3c}, based on explicit disparity estimation and image warping.

Several learned methods have been proposed for lossy stereo compression. Liu et al.\cite{Liu2019DSICDS} proposed the DSIC method, which uses a conditional entropy model where skip modules feed disparity warped features from the encoded first image into the second. Deng et al.\cite{deng2021hesic} proposed the HESIC method, in which the second image is warped by an estimated homography, and only the residual is encoded. In addition, a context-based entropy model and a final quality enhancement module are used to reduce the bitrate and increase the reconstruction quality. This has been simplified in the SASIC method proposed by W\"odlinger et al. \cite{wodlinger2022sasic}, where the transformation between the images in the stereo pair is approximated by a channel-wise horizontal shift and only the residual for the second image is encoded. This is further enhanced by using stereo attention between the two images in the common decoder. Cross-attention in the decoder is also used in the recently proposed distributed multi-view method LDMIC by Zhang et al. \cite{zhang2023ldmic}. Contrary to our method, they employ global encoder-to-decoder cross-attention and a single-image autoregressive entropy model. Mital et al. \cite{mital2023neural} propose a similar approach for the special case of distributed source coding where a correlated image is available during decoding.

Unlike HESIC or SASIC, the proposed method does not seek a parametric transformation between the stereo images and does not use residual coding. Different from LDMIC, our stereo attention module only attends to the epipolar line, resulting in a significant runtime reduction without any penalty in the performance. Another difference to LDMIC is that our method does not use autoregressive context and is therefore much faster in decoding. Instead, we condition the entropy model of the right image on the left image, effectively using the left image as a context without a significant runtime penalty.

%%%%%%%%%%%%%%%%%%%%%%%%%%%%%%
%%%%%%%%%%% METHOD %%%%%%%%%%%
%%%%%%%%%%%%%%%%%%%%%%%%%%%%%%
\section{Method}

The proposed method follows the common structure~\cite{balle2018variational} consisting of the main autoencoder and hyperprior, to which we add two non-autoregressive context modules; see overview in Fig.~\ref{fig: overview}. In the main branch, consisting of the encoder $E$ and decoder $D$, the input image pair $(\bm{x}_l, \bm{x}_r)$ is transformed to the latent representation $(\bm{{y}}_l, \bm{{y}}_r)$ and quantized to discrete tensors $(\bm{\hat{y}}_l, \bm{\hat{y}}_r)$ which constitutes the bitstream. The decoder $D$ reconstructs the output images $(\bm{\hat x}_l, \bm{\hat x}_r)$. In the hyperprior branch the hyper-encoder $h_E$ transforms the latents to $(\bm{{z}}_l, \bm{{z}}_r)$ which again undergo quantization to $(\bm{\hat{z}}_l, \bm{\hat{z}}_r)$ and are stored in the bitstream as side-information. They are then used by the hyper-decoder $h_D$ to estimate the entropy parameters of the latents $(\bm{\hat{y}}_r, \bm{\hat{y}}_r)$. All these modules jointly process the left and right images in parallel.

We add two non-autoregressive {\it stereo context modules} $c_y$ and $c_z$ to the right image stream, which aid in estimating the entropy parameters of the right image latents $\bm{\hat{y}}_r$ and hyperlatents $\bm{\hat{z}}_r$, respectively, using the information already available from the left side. The left and right image stream are further connected via the proposed {\it Stereo Cross Attention} (SCA) modules (see Section~\ref{sec: sca}), which are included in all the modules that connect both streams~--~in encoder/decoder and hyper-encoder/hyper-decoder~--~and also in the stereo context $c_y$.

The resulting method can be trained end-to-end with the rate-distortion loss (see Sec.~\ref{sec: loss}) on any dataset of stereo image pairs. Unlike other recent methods \cite{deng2021hesic, zhang2023ldmic}, the proposed method does not include any autoregressive components, which allows for fast encoding and decoding (see Sec.~\ref{sec: complexity}).

\subsection{Encoding/Decoding and Quantization}
The encoder $E$ and decoder $D$ each consist of four convolutional layers with three down/upsampling steps for left and right each, one SCA module, and PReLU \cite{he2015delving} activation functions. The encoder and decoder of the hyperprior similarly employ three convolutional layers with two down/up-sampling steps, one SCA module, and PReLU activation functions. The initial convolutions in the encoder $E$ share weights between left and right streams. Network diagrams for each module can be found in the supplementary material. 
Each quantization operation applies integer rounding to the mean-subtracted input. For example for the left latent:
\begin{equation}
    \bm{\hat{y}}_l = \operatorname{round}(\bm{y}_l - \bm{\mu}_l) + \bm{\mu}_l,  
\end{equation}
where $\bm{\mu}_l$ is the estimated mean of the distribution of $\bm{y}_l$. Analogously for $\bm{y}_r, \bm{z}_l$, and $ \bm{z}_r$.

\subsection{Stereo Cross Attention Module}
\label{sec: sca}
\begin{figure}[t]
	\begin{center}
			\includegraphics[width=0.5\textwidth]{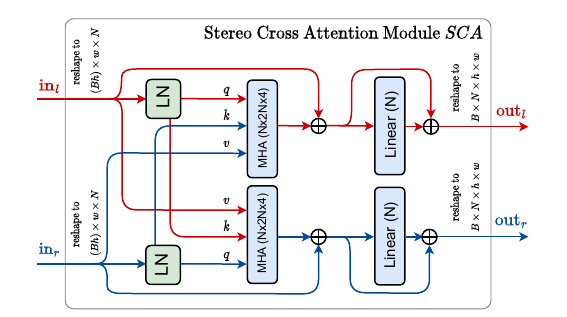}
	\end{center}
	\caption{\label{fig: sca}%
	The proposed Stereo Cross Attention (SCA) module. The left and right streams are colored red and blue, respectively. LN denotes layer norm and MHA denotes multi-head attention block with arguments (output dimension $\times$ embedding dimension $\times$ heads). The streams denoted $q$, $k$, and $v$ refer to the standard query, key and value terminology.}
\end{figure}

We propose a new Stereo Cross Attention (SCA) module to facilitate the flow of non-local information between the left and right image compression stream. It performs cross attention between the corresponding epipolar lines~--~for each location in one image the attention domain is the corresponding horizontal row in the other image. By restricting the attention only to horizontal lines (under the assumption that the input images are rectified) we circumvent the issue of the quadratic memory complexity of vanilla attention and can process all rows in parallel. The resulting method still has quadratic complexity but only in the width rather than the total pixel count $\mathcal{O}(w^2h)$. The structure of the SCA module is shown in Fig.~\ref{fig: sca}. Layer norm is applied only to queries and keys (not values, which constitute the final output). In the Multi-Head Attention (MHA) block, we use 1D convolutions with a kernel size of $3$ instead of linear embeddings. We also tried different variants of positional encoding \cite{shaw2018self, vaswani2017attention} but found no impact on overall performance.
The SCA module is included in all of the submodules that connect both streams~--~$E, D, h_E, h_D$, and also in the stereo context $c_y$; see Fig.~\ref{fig: overview}. In $E$ and $h_E$, the SCA module is applied after all downsampling layers and before the final convolutional layer. In $D$ and $h_D$, the module is applied after the initial upsampling layer. Additional details can be found in the supplementary material.

\subsection{Entropy model}
\label{sec: sc}
Following the hyperprior structure from \cite{balle2018variational} we employ a pair of hyperlatents $\bm{\hat{z}}_l, \bm{\hat{z}}_r$ as side information to aid in estimating the entropy parameters of the main latents $\bm{\hat{y}}_l, \bm{\hat{y}}_r$. In the following paragraphs, we write tensors (non-scalars) in bold, use $\bm{\theta}_{(\cdot)}$ for entropy parameters and $\bm{\phi}_{(\cdot)}$ for other learnable or predicted parameters.

The distribution of the left hyperlatent $\bm{\hat{z}}_l$ is modeled by a channel-wise Laplacian distribution $\text{Lap}_{\bm{\mu}, \bm{b}}$ with parameters $\bm{\theta}_z^l := (\bm{\mu}_z^l, \bm{b}_z^l)$ for each channel of $\bm{\hat{z}}_l$ that are learned during training and fixed afterwards. The distribution of the right hyperlatent is modeled by a factorized Laplace distribution with parameters $\bm{\theta}_z^r := (\bm{\mu}_z^r, \bm{b}_z^r)$ for each pixel. These are predicted adaptively for each input. Likewise, the distribution of the main latents $\bm{\hat{y}}_{l/r}$ is also modeled by a factorized Laplace distributions with parameters $(\bm{\theta}_y^l,\bm{\theta}_y^r)$.

\begin{figure}[t]
	\centering
	\setlength{\tabcolsep}{1pt}
	\begin{tabular}{l}
        \hspace{.01\textwidth}\includegraphics[width=.49\textwidth]{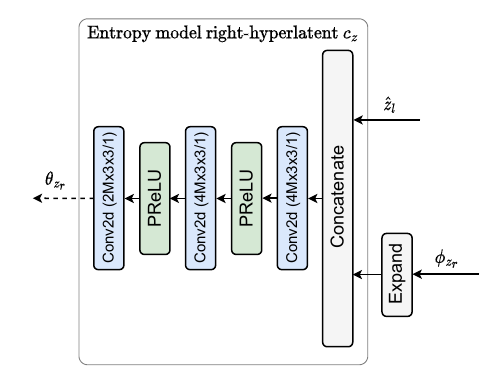}\\
    	\includegraphics[width=.45\textwidth]{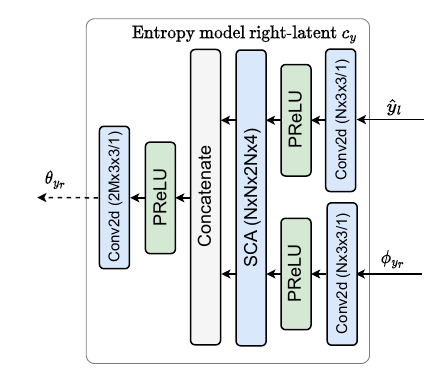}
	\end{tabular}
	\caption{The context modules $c_z$ and $c_y$. Top: $c_z$ predicts the entropy parameters of the right hyperlatent $\bm{\hat{z}}_r$ from the left hyperlatent $\bm{\hat{z}}_l$ and a set of learned parameters $\bm{\phi}_{\bm{z}_r}$. Bottom: $c_y$ predicts the entropy parameters of the right latent $\bm{\hat{y}}_r$ from the left latent $\bm{\hat{y}}_l$ and partial output of the hyper-decoder $h_D$ denoted $\bm{\phi}_{\bm{y}_r}$. In our experiments, we set $N = 192$ and $M = 48$. The arguments after convolutions denote (output dimension $\times$ kernel / stride).}
	\label{fig: context modules}
\end{figure}

To reduce the bitrate we condition the right image entropy model on information from the left stream. To this end, we include two stereo context modules, $c_y$ and $c_z$; see Fig.~\ref{fig: context modules}.

The left hyperlatent entropy parameters are learned. The right hyperlatent entropy parameters $\bm{\theta}_z^r$ are predicted by $c_z$ from $\bm{\hat{z}}_l$ and a set of fixed (learnable) parameters $\bm{\phi}_{\bm{z}_r}$:
\begin{equation}
    \bm{\theta}_z^r = c_z(\bm{\hat{z}}_l,\bm{\phi}_{\bm{z}_r}).
\end{equation}
During encoding (decoding), $\bm{\hat{z}}_l$ is encoded (decoded) first using its fixed entropy model and then used to encode (decode) $\bm{\hat{z}}_r$ using entropy parameters predicted by $c_z$.

The parameters $\bm{\theta}_y^l$ of the distribution of the left latent $\bm{\hat{y}}_l$ are predicted from both hyperlatents $\bm{\hat{z}}_l, \bm{\hat{z}}_r$ by the hyperprior decoder $h_D$. Similarly to the previous case, we use the decoded left latent $\bm{\hat{y}}_l$ to more accurately estimate the entropy parameters of the right latent. To this end we include the context module $c_y$ which predicts the right latent entropy parameters
\begin{equation}
    \bm{\theta}_y^r = c_y(\bm{\hat{y}}_l, \bm{\phi}_{\bm{y}_r})
\end{equation}
from the already decoded left latent and the second output $\bm{\phi}_{\bm{y}_r}$ of the hyper-decoder $h_D$.

\subsection{Loss Function}
\label{sec: loss}
In training we optimize the rate-distortion loss
\begin{equation}
    \mathcal{L} = \mathcal{R} + \lambda \mathcal{D},
\end{equation}
where $\mathcal{R}$ denotes the rate and $\mathcal{D}$ the distortion loss term; $\lambda \in \mathbb{R}$ is a trade-off parameters that determines the average bitrate of the trained model. The distortion loss term is the expectation of the mean-squared-errors
\begin{equation}
    \mathcal{D}(\bm{x}_l, \bm{x}_r) = \mathbb{E}_{\bm{x}_l, \bm{x}_r \sim p_{\bm{x}}} \Big[ \lVert \bm{x}_l - \bm{\hat{x}}_l \rVert_2^2 + \lVert \bm{x}_r - \bm{\hat{x}}_r \rVert_2^2 \Big].
\end{equation}

To estimate the rate, we compute the cross entropy between the predicted distribution of our entropy model and the true distribution of the latents/hyperlatents. The total rate loss is then the sum of the rates of the latents and the hyperlatents:
\begin{align}
    \begin{split}
        \mathcal{R} = \mathbb{E}_{\bm{x}_l, \bm{x}_r \sim p_{\bm x}} \big[
        &-\log_2 p( \bm{\hat{z}}_l\mid\bm{\theta}_z^l)\\
        &-\log_2 p( \bm{\hat{z}}_r\mid \bm{\phi}_{c_z}, \bm{\phi}_{\bm{z}_r}, \bm{\hat{z}}_l)\\
        &-\log_2 p( \bm{\hat{y}}_l\mid \bm{\phi}_{hd}, \bm{\hat{z}}_r, \bm{\hat{z}}_l)\\
        &-\log_2 p( \bm{\hat{y}}_r\mid \bm{\phi}_{c_y}, \bm{\phi}_{hd}, \bm{\hat{y}}_l, \bm{\hat{z}}_r, \bm{\hat{z}}_l) \big],
    \end{split}
\end{align}
where $\bm{\phi}_{hd}, \bm{\phi}_{c_y}, \bm{\phi}_{c_z}$ denote the parameters of the hyperprior decoder and our proposed stereo context modules $c_y$ and $c_z$, respectively, and $p(\ldots)$ are the Laplace distributions specified in the preceding section \ref{sec: sc}.

Since quantization has zero derivative almost everywhere it needs to be replaced by some proxy expression during training. As in \cite{balle2017end} we approximate quantization with additive uniform noise for the rate loss (similarly for $\bm{y}_r, \bm{z}_l$ and $\bm{z}_r$)
\begin{equation}
    \bm{\Tilde{y}}_l = \bm{y}_l + \mathcal{U}(-0.5, 0.5).
\end{equation}
Following Minnen et al. \cite{minnen2020channel}, we employ straight-through-estimation quantization during training for the distortion loss.

%%%%%%%%%%%%%%%%%%%%%%%%%%%%%%
%%%%%%%% EXPERIMENTS %%%%%%%%%
%%%%%%%%%%%%%%%%%%%%%%%%%%%%%%
\section{Experiments}
We give a brief overview of the datasets on which we evaluate our method, followed by implementation details. We then show rate-distortion curves, an analysis of encoding/decoding times, and conclude the section with an ablation study.

\subsection{Experimental Setup}

\paragraph{Datasets:} We evaluate our method on two popular stereo image datasets, Cityscapes \cite{cordts2016cityscapes} and InStereo2k \cite{bao2020instereo2k}. The datasets capture two different settings for stereo imagery. Cityscapes contains stereo pairs of driving scenes with large variations of disparities in a single image pair. 
The InStereo2k dataset, on the other hand, contains only indoor scenes of assortments of objects closer to the camera. The Cityscapes dataset contains 5000 stereo image pairs of urban street scenes taken while driving in German cities. The images have a resolution of $2048 \times 1024$ and are divided into 2975 training, 500 validation, and 1525 test image pairs. Following conventions \cite{zhang2023ldmic, wodlinger2022sasic} we crop 64, 256 and 128 pixels from the top, bottom and sides respectively to remove car parts and artefacts from the rectification process. The InStereo2k dataset contains 2060 stereo images of indoor scenes. The images have a resolution of $1080 \times 860$ and are divided into 2010 training and 50 test image pairs. We crop the images symmetrically so that the height and width are multiples of 32.

\paragraph{Implementation Details:} In all experiments we set the number of channels in the encoding and decoding modules to $N = 192$ and the number of latent channels to $M = 48$. We use 4 headed attention blocks with an embedding dimension of $386$. To generate the rate-distortion curves, we train our method for 450 epochs on cityscapes and for 650 epochs on InStereo2k ($\sim$1.3M steps each). We use the Adam optimizer with an initial learning rate of $10^{-4}$ for a batch size of $1$ and reduce the learning rate to $10^{-5}$ after 1M steps. We vary $\lambda$ depending on the targeted bitrate. We train on random crops of size $256 \times 1024$ and evaluate on full-resolution images (except for dataset-specific crops as specified above).

\paragraph{Codecs:} We compare our method with an extensive list of conventional and learned codecs. The codecs can be broadly grouped into four categories: single-image compression (BPG), video compression (HEVC), multi-view compression (MV-HEVC, LDMIC), and stereo image compression (HESIC/HESIC+, DSIC, SASIC, ECSIC (ours)). BPG \cite{bpg} is applied to each frame independently without chroma subsampling. The video codecs HEVC \cite{hevc} and VVC \cite{vvc} are applied to a stereo image pair as two-frame video sequence with chroma subsampling disabled as it would unnecessarily degrade the PSNR score. We use the official reference implementation\footnote{\url{https://vcgit.hhi.fraunhofer.de/jvet/HM}} with \verb|main_444_12| profile (YUV444 12bit) for HEVC. For VVC we report the values from Zhang et al. \cite{zhang2023ldmic} where the \verb|lowdelay_p| configuration as well as YUV444 input format is used. MV-HEVC\footnote{\url{http://hevc.info/mvhevc}} \cite{mvhevc} is used in the two-view intra-mode configuration (unfortunately, it only supports 4:2:0 chroma mode, resulting in worse PSNR scores at higher bitrates). We also report scores for the learned stereo compression methods DSIC \cite{Liu2019DSICDS}, HESIC+ \cite{deng2021hesic}, and SASIC \cite{wodlinger2022sasic} from the respective papers. For LDMIC \cite{zhang2023ldmic} we show the reported scores for the full LDMIC method, which includes an autoregressive context model, and a smaller version LDMIC (fast) without the autoregressive components.

\subsection{Results}
\begin{figure*}
    \centering
    \hfill\includegraphics[width=.5\linewidth]{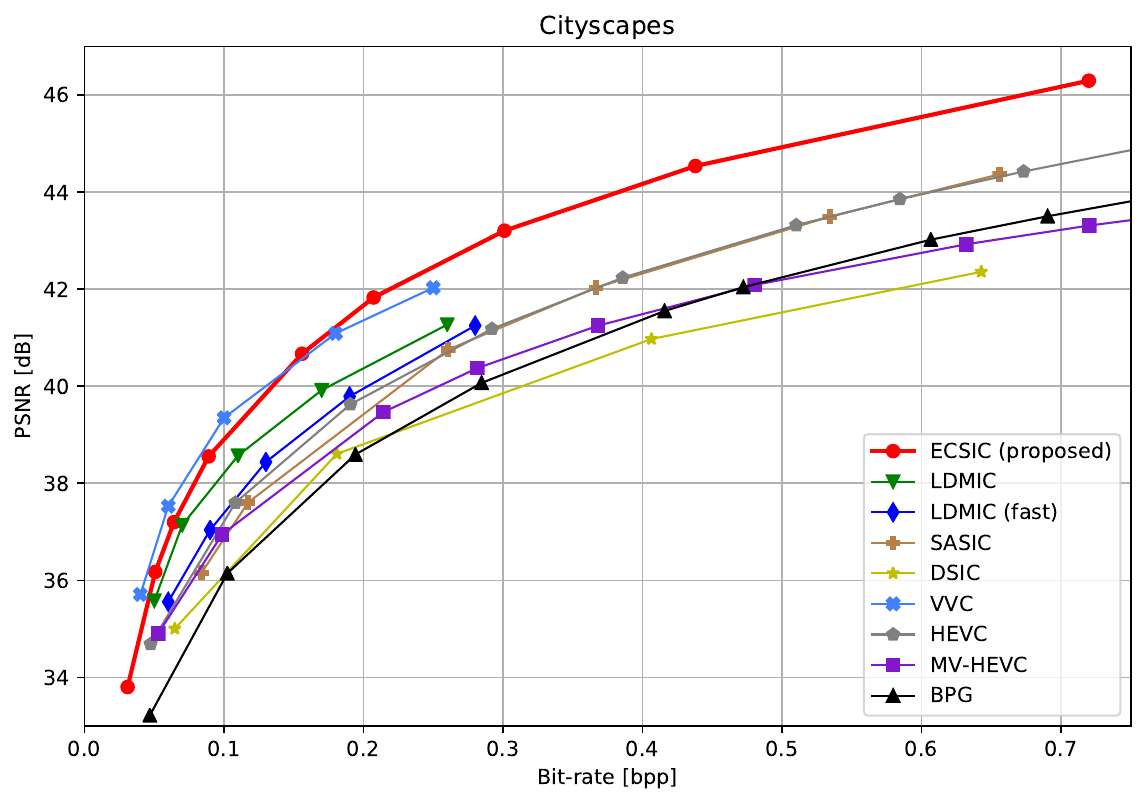}\hfill%
    \includegraphics[width=.5\linewidth]{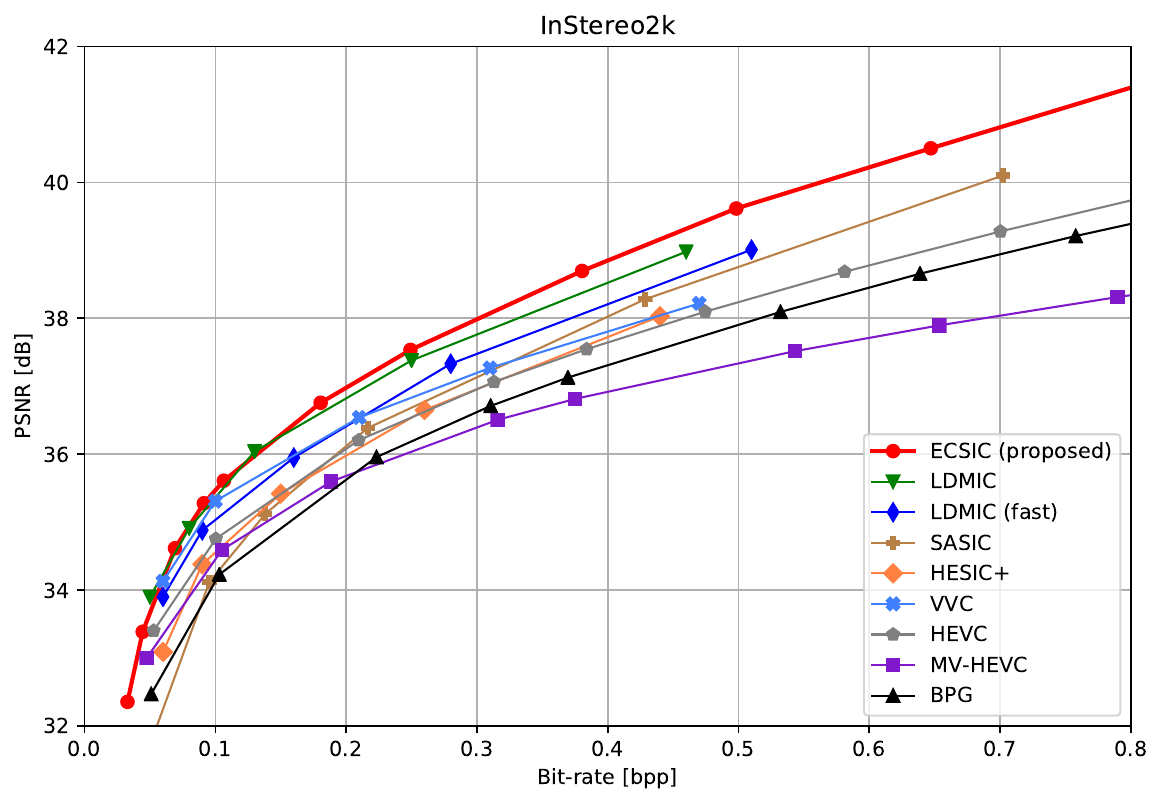}\hfill\null%
    \caption{Rate-distortion curves of our method and other codecs on Cityscapes (left) and InStereo2K (right) datasets measured by PSNR.}
    \label{fig: rd curves}
\end{figure*}

The Peak Signal to Noise Ratio (PSNR) rate-distortion curves can be seen in Fig.~\ref{fig: rd curves}. We also report the Bjøntegaard Delta bitrate (BD-Rate) and BD-PSNR scores~\cite{bjontegaard2001} for each codec w.r.t. BPG for Cityscapes and InStereo2K in Tab.~\ref{tab: cityscapes bd scores} and Tab.~\ref{tab: instereo bd scores}, respectively. On InStereo2k, the proposed method outperforms all other codecs tested. On Cityscapes, Our method performs worse than VVC for low bpp (bits per pixel) ($<0.15$) but is the first learned method to outperform VVC in PSNR for $\text{bpp} > 0.15$. All other codecs are outperformed by our method. The second best learned method LDMIC relies on autoregressive entropy modelling in the default version which renders it much slower than our method (see Sec.~\ref{sec: complexity}). The fast variant of LDMIC without autoregressive context performs much worse with BD-Rate w.r.t. to ECSIC of $24.35\%$ for InStereo2k and $49.23\%$ for Cityscapes. 

Our method also outperforms the other learned SIC models DSIC, HESIC+ and SASIC. Both HESIC+ and SASIC rely on explicit warping to remove spatial redundancies. HESIC+ uses a homography warping and SASIC applies a channel-wise shift to the latents. Interestingly, SASIC also uses a variant of cross attention in its decoder. The performance gap between SASIC and the proposed method shows the effectiveness of the proposed SCA and stereo context modules.

\begin{figure*}
	\centering
	\includegraphics[width=\linewidth]{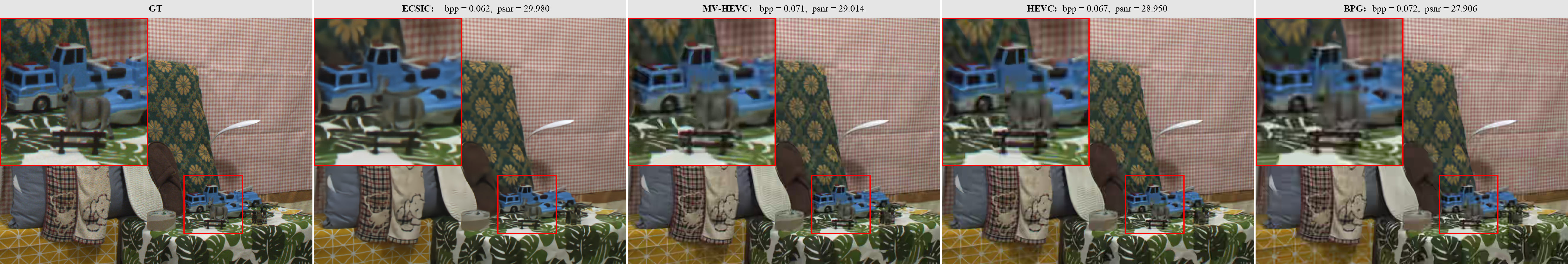}
	\caption{A qualitative comparison of our method against other codecs on an image from the InStereo2K test set. The zoomed-in region is upscaled with nearest neighbour upsampling.}
	\label{fig: 001860}
\end{figure*}

We provide a qualitative comparison of our method against a selected list of codecs on an image from the InStereo2k dataset in Fig.~\ref{fig: 001860}. Additional qualitative comparisons for samples from both the Cityscapes and InStereo2k datasets as well as MS-SSIM rate distortion curves can be found in the supplementary material.


\begin{table}
    \caption{Relative quality difference (PSNR gain at the same bitrate; higher is better) and bitrate difference (bitrate gain for the same PSNR; lower is better) of the benchmarked methods on Cityscapes w.r.t. BPG. Best results in bold and second best underlined.}
    \label{tab: cityscapes bd scores}
    \centering
    \vspace{0.3cm}
    \begin{tabular}{lrr}
        \toprule
        Method & BD-PSNR [dB]$\uparrow$ & BD-Rate [\%]$\downarrow$ \\
        \hline
        % These numbers are with respect to BPG vvvv
        ECSIC (proposed)  & \underline{2.86}  & \underline{-51.90} \\
        LDMIC           & 2.07         & -42.20 \\
        LDMIC (fast)    & 1.35         & -29.66 \\
        SASIC           & 0.98         & -22.40 \\
        DSIC            & 0.07         & -3.35 \\
        VVC             & \bf{3.12}    & \bf{-56.24} \\
        HEVC            & 1.14         & -25.78 \\
        MV-HEVC         & 0.41         & -10.07 \\
        BPG             & 0.0         & -0.0 \\
        \toprule
    \end{tabular}
\end{table}
\begin{table}
    \caption{Relative quality difference (PSNR gain at the same bitrate; higher is better) and bitrate difference (bitrate gain for the same PSNR; lower is better) of the benchmarked methods on InStereo2K w.r.t. BPG. Best results in bold and second best underlined.}
    \label{tab: instereo bd scores}
    \centering
    \vspace{0.3cm}
    \begin{tabular}{lrr}
        \toprule
        Method & BD-PSNR [dB]$\uparrow$ & BD-Rate [\%]$\downarrow$ \\
        \hline
        % These numbers are with respect to BPG vvvv
        ECSIC (proposed)& \bf{1.57}    & \bf{-42.08} \\
        LDMIC           & \underline{1.26}          & \underline{-41.03} \\
        LDMIC (fast)    & 0.87         & -30.40 \\
        SASIC           & 0.38         & -15.43 \\
        HESIC+          & 0.37         & -14.90 \\
        VVC             & 0.86         & -31.02 \\
        HEVC            & 0.45         & -15.09 \\
        MV-HEVC         & 0.19         & -4.96 \\
        BPG             & 0.0           & -0.0 \\
        \toprule
    \end{tabular}
\end{table}

\subsection{Ablation Study}
\begin{figure*}
    \centering
    \hfill\includegraphics[width=.5\linewidth]{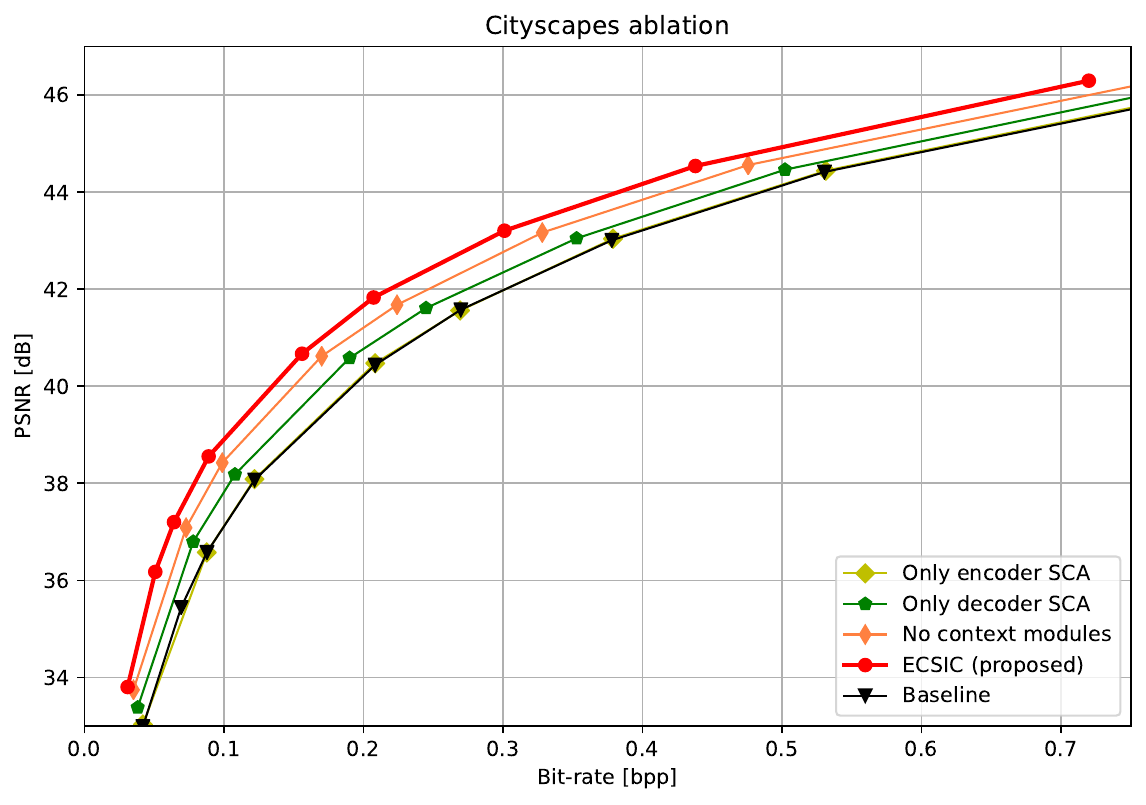}\hfill%
    \includegraphics[width=.5\linewidth]{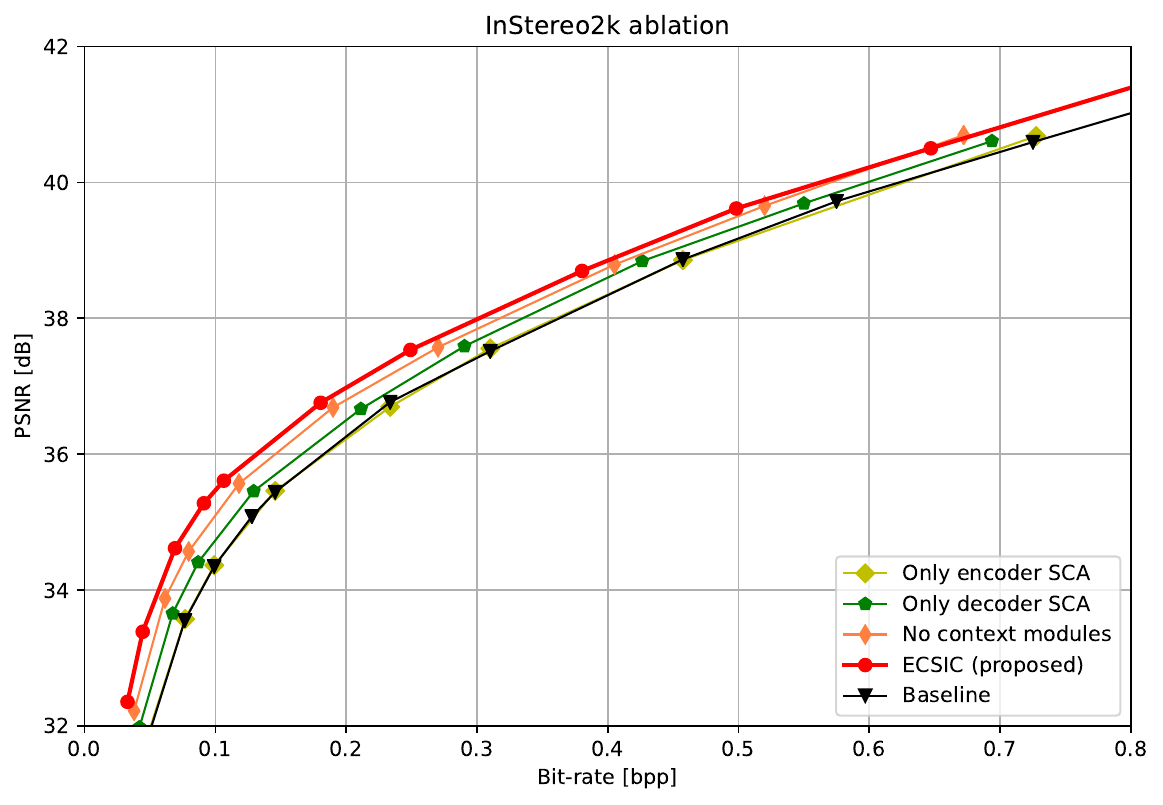}\hfill\null%
    \caption{Rate-distortion curves for our method with varying modifications for Cityscapes (left column) and InStereo2K (right column) measured by PSNR.}
    \label{fig: rd curves ablation}
\end{figure*}

To assess the impact of individual components/additions in our method compared to a single image compression baseline, we generate rate-distortion curves on Cityscapes and InStereo2k for various architectural modifications of our method. The resulting rate-distortion curves are shown in Fig.~\ref{fig: rd curves ablation}.

\paragraph{Baseline:} To assess the impact of individual components on stereo compression performance, we construct a baseline method by stripping the ECSIC model of both context modules $c_y$ and $c_z$ (the entropy modelling for left and right is independent of each other) and removing each SCA module in the remaining architecture (including the corresponding activation functions of each SCA layer). The result is two separate models that independently compress the left and right images of a stereo-image pair.

\paragraph{ECSIC (proposed):} The proposed method with all additions as shown in Fig.~\ref{fig: overview}. The largest gains over the baseline are at low bit rates. For example, for Cityscapes, the BD-rate limited to low PSNR ($34 - 38$dB) shows a bitrate saving of $37.0\%$, while in the high PSNR range ($44 - 46$dB) the difference is reduced to $19.0\%$. The maximum asymptotic theoretical bitrate saving that can be achieved is $50.0\%$, which corresponds to compressing a stereo image pair with the bitrate of a single image. In reality, the optimum is even lower due to occlusions and non-overlapping fields of view in the stereo pair.

\paragraph{Only encoder SCA:} We extend the baseline method by adding a single SCA layer to the encoder $E$. The resulting model shows no significant improvement over the baseline method. However, we have found that adding SCA modules in the encoder does improve performance if corresponding SCA modules are present in the decoders.

\paragraph{Only decoder SCA:} We extend the baseline method by adding a single SCA layer in the decoder $D$. Contrary to adding SCA only in the encoder, adding a single SCA layer in the decoder already gives an improvement of $11.7 \%$ over the baseline method on Cityscapes.

\paragraph{No context modules:} Removing the context modules $c_y$ and $c_z$ results in a $12.5\%$ rate reduction compared to the full ECSIC model.
\\ \\
Our comparison shows that both, the proposed context modules $c_y$ and $c_z$ as well as the proposed SCA modules enable better compression of stereo images when compared to a single image compression baseline model that compresses both images independently. Furthermore, our experiments suggest that the SCA module works best in the decoding parts of the model. However, we found that SCA modules in the encoding part of the model in conjunction with SCA modules in the corresponding decoding parts of the model leads to the best overall performance. We also conducted experiments with different variants of positional encoding \cite{shaw2018self, vaswani2017attention} but found no impact on compression performance.

\subsection{Coding Complexity}
\label{sec: complexity}
\begin{figure}
    \centering
    \setlength{\tabcolsep}{1pt}
    \begin{tabular}{c}
        \includegraphics[width=\linewidth]{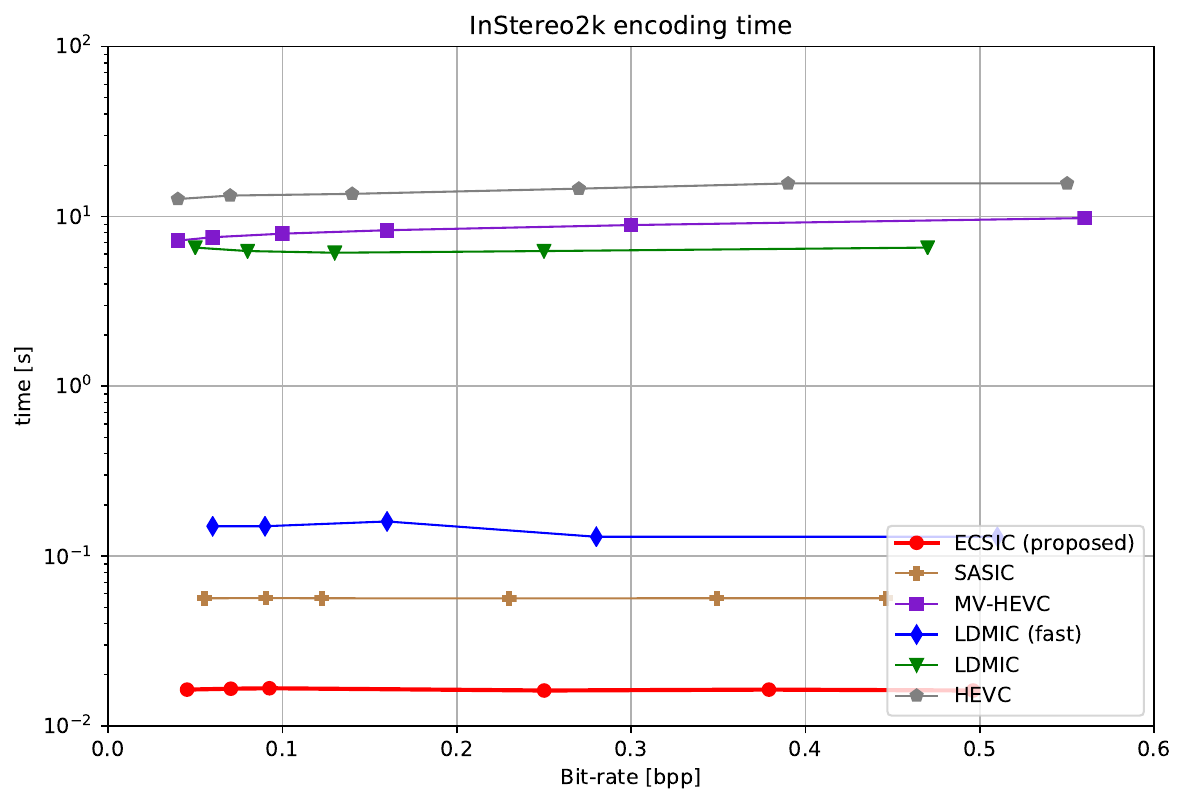}\\
        \includegraphics[width=\linewidth]{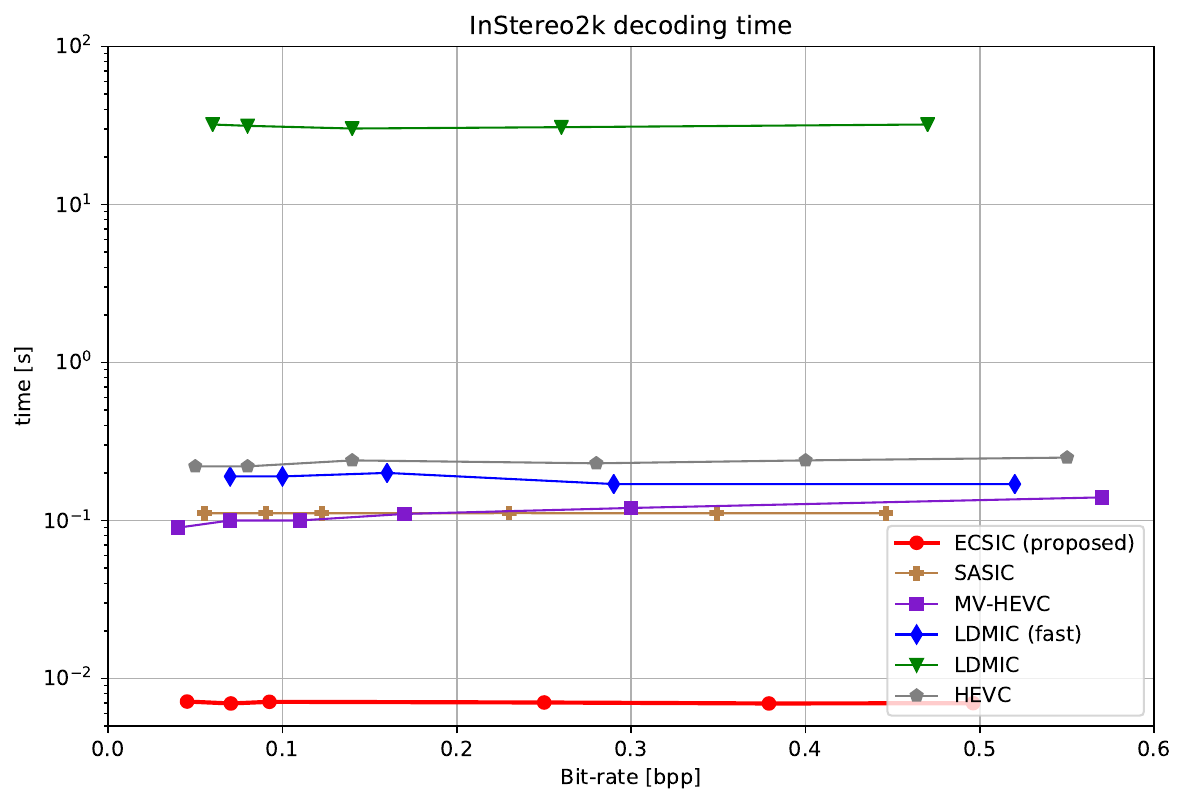}
    \end{tabular}
    \caption{Encoding and decoding times of our method against other codecs on a logarithmic scale. The reported times are averages over the InStereo2k test set. All learned methods run on a GPU.}
    \label{fig: timing curves}
\end{figure}

Fig.~\ref{fig: timing curves} depicts the average encoding and decoding times of our method against other methods on the InStereo2k dataset (i.e. the images are already rectified). The conventional methods BPG, HEVC and MV-HEVC were evaluated on an Intel Xeon Gold 6230R processor with a single core (times are taken from Zhang et al. times~\cite{zhang2023ldmic}). For LDMIC we show their reported encoding and decoding times~\cite{zhang2023ldmic} (measured on an NVIDIA RTX 3090 GP U). For SASIC and our proposed ECSIC model we measure encoding and decoding times on an NVIDIA RTX 3090 GPU. To make the comparison with conventional methods running on CPU fair, we also include the time needed to load the image and move it on the same GPU as the model in the encoding time. On our machine, this takes on average about $5$ms which, for ECSIC, is slightly less than one third of the total reported encoding time. The proposed method shows low encoding and decoding times, beating all other methods in this benchmark.

\subsection{Limitations}
Our evaluation is, as is typical in the field of stereo image compression, limited to particular available datasets where training and test sets are from the same distribution. It is expected that the resulting method will not be able to generalize to different camera setups if no additional care is taken during training (e.g., larger and more diverse training sets with varying camera setups and potentially concurrently increasing the model size). Furthermore, since our SCA module restricts the cross-attention operation essentially to the epipolar line, the input pairs must be rectified first. Finally, our method relies on a GPU or some other neural hardware accelerator to achieve the reported encoding/decoding times.

%%%%%%%%%%%%%%%%%%%%%%%%%%%%%%
%%%%%%%%% CONCLUSION %%%%%%%%%
%%%%%%%%%%%%%%%%%%%%%%%%%%%%%%
\section{Conclusion}
In this work, we have presented ECSIC, a novel method for stereo image compression. Our method consists of a convolutional neural network augmented with stereo attention modules that enable the network to compress both images jointly by exploiting the mutual information between them. Additionally, we proposed two stereo context modules for better entropy modelling of stereo images and showed in an ablation study the effectiveness of the proposed methods. The resulting model is fast in both encoding and decoding and outperforms all other learned compression methods on the two stereo image benchmark datasets Cityscapes and InStereo2k.

%%%%%%%%%%%%%%%%%%%%%%%%%%%%%%
%%%%%% ACKNOWLEDGEMENTS %%%%%%
%%%%%%%%%%%%%%%%%%%%%%%%%%%%%%
\section{Acknowledgements}
This project has received funding from the European Union's Horizon 2020 research and innovation program under grant agreement No 965502.
\FloatBarrier 

\newpage
{\small
\bibliographystyle{ieee_fullname}
\bibliography{egbib}
}

% \newpage
%%%%%%%%%%%%%%%%%%%%%%%%%%%%%%
%%%%%% ACKNOWLEDGEMENTS %%%%%%
%%%%%%%%%%%%%%%%%%%%%%%%%%%%%%
\section{Appendix}

We provide additional results for our ablation study in Subsection~\ref{subsec: ablation addon}, MS-SSIM\cite{msssim} rate-distortion curves in Subsection~\ref{subsec: ms-ssim}, a detailed description of the layer structure of our model in Subsection~\ref{subsec: submodules}, and finally additional qualitative examples in Subsection~\ref{subsec: qualitative results}.

\subsection{Additional Ablation Study Results}
\label{subsec: ablation addon}

\begin{figure}
    \centering
    \begin{tabular}{c}
         \includegraphics[width=.5\textwidth]{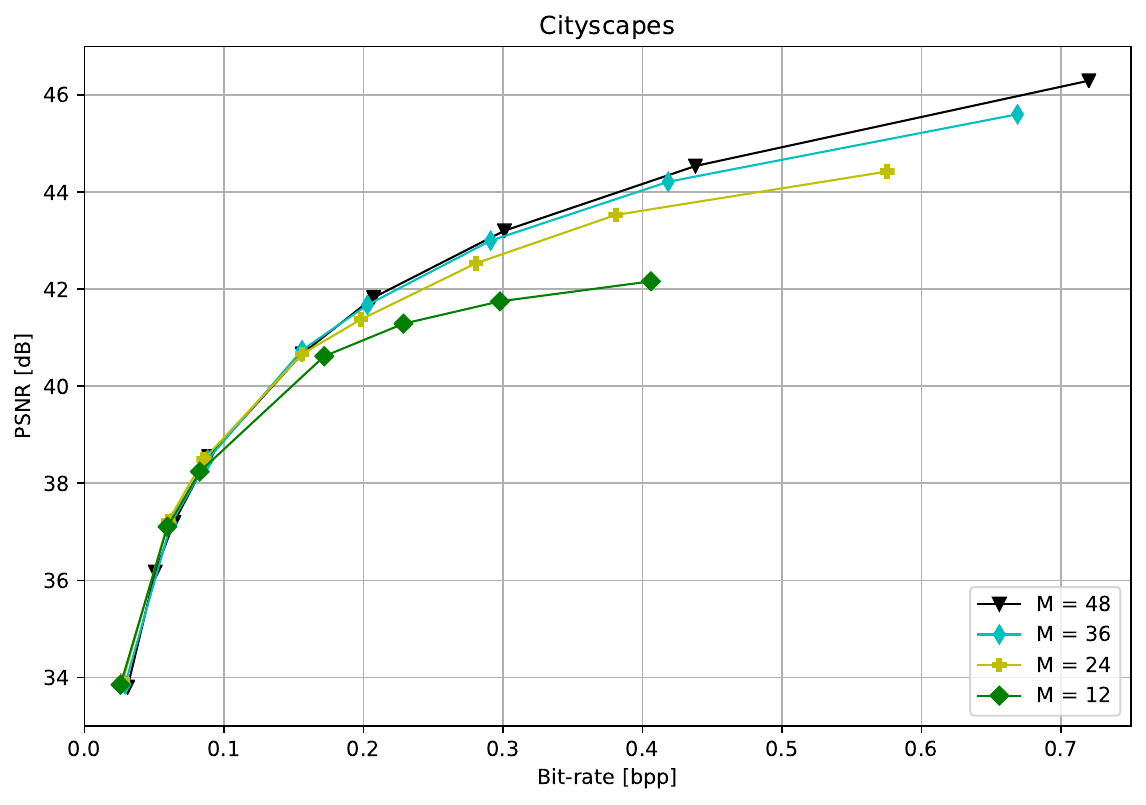}\\
         \includegraphics[width=.5\textwidth]{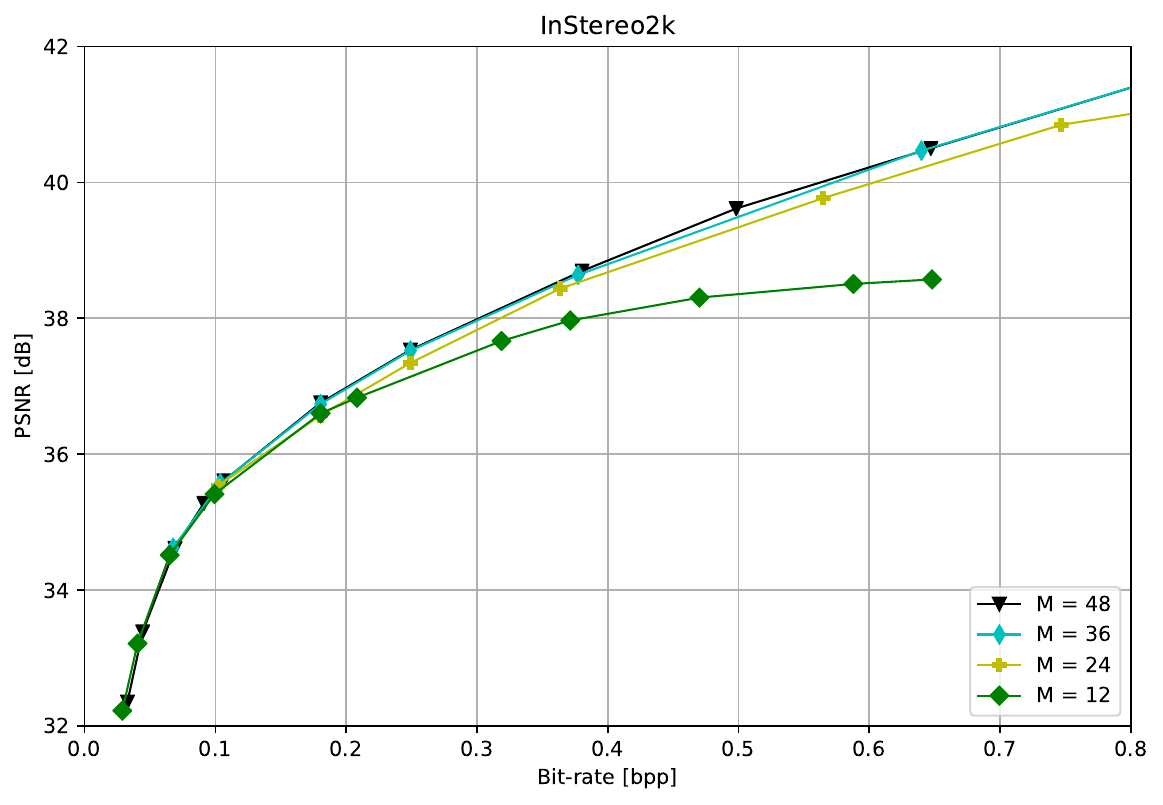}
    \end{tabular}
    % \hfill\includegraphics[width=.5\linewidth]{figures/cityscapes_m_variation.pdf}\hfill%
    % \includegraphics[width=.5\linewidth]{figures/instereo_m_variation.pdf}\hfill\null%
    \caption{Effects of varying values for the number of latent channels $M$.}
    \label{fig: m variation}
\end{figure}

\paragraph{Number of latent channels:} Figure~\ref{fig: m variation} shows the PSNR rate distortion curves for different values of the parameter $M$, i.e., the number of channels in the latents $\bm{\hat{y}}_l, \bm{\hat{y}}_r$ and hyperlatents $\bm{\hat{z}}_l, \bm{\hat{z}}_r$. At high bitrates the quality of the reconstruction plateaus, for lower values for $M$, for both datasets, indicating that at higher bitrates small values for $M$ are an overly restrictive bottleneck. At very low bitrates on the Cityscapes dataset, the $M = 12$ model outperforms $M = 48$ with a bitrate reduction of $~17\%$ for comparable PSNR, but the absolute difference is minimal. For medium to high bitrates, however, the $M = 48$ model performs much better. We therefore set $M = 48$ for all other experiments as it gives the best overall performance. Values for $M$ larger than $48$ did not show any further improvement in the range of bitrates tested.

\paragraph{BD-Scores:} We provide quantitative measurements of the performance differences between different experiments from our ablation study in the main paper. We show the Bjøntegaard Delta bitrate~(BD-Rate)\cite{bjontegaard2001} and BD-PSNR scores relative to the single image baseline in Table~\ref{tab: cityscapes bd table} for Cityscapes and Table~\ref{tab: instereo bd table} for InStereo2k. The description of the experiments can be found in Section 4.3 of the main paper. We also provide BD rate scores for higher quality (higher bitrates) and lower quality (lower bitrates). Our method shows consistent improvements of $37\%$\footnote{The maximum asymptotic theoretical bitrate saving is $50.0\%$, which is equivalent to compressing a stereo image pair at the bitrate of a single image. Due to occlusions and non-overlapping fields of view, the true optimum is even lower.} in the low bitrate range for both datasets. At high bitrates, the rate savings decrease to $19\%$ for Cityscapes and $12.6\%$ for InStereo2k. 
\begin{table*}
    \caption{Relative quality difference (PSNR gain at the same bitrate; higher is better) and bitrate difference (bitrate gain for the same PSNR; lower is better) of the benchmarked methods on Cityscapes w.r.t. the baseline model. We also report the BD-Rate restricted to a low PSNR region ($34-38$dB) and high PSNR region ($44-46$dB).}
    \label{tab: cityscapes bd table}
    \centering
    \vspace{0.3cm}
    \begin{tabular}{lcccc}
        \toprule
        Method & BD-PSNR [dB]$\uparrow$ & BD-Rate [\%]$\downarrow$ & \tb{BD-Rate [\%]$\downarrow$}{low PSNR} &\tb{BD-Rate [\%]$\downarrow$}{high PSNR} \\
        \hline
        % These numbers are with respect to BPG vvvv
        ECSIC (proposed)    & \bf{1.49}     & \bf{-30.18}    & \bf{-36.96}   & \bf{-19.02} \\
        No context modules  & 1.02          & -21.45         & -26.95        & -12.74 \\
        Only decoder SCA    & 0.54          & -11.72         & -15.57        & -6.05 \\
        Only encoder SCA    & 0.02          & -0.40          & -0.36         & -0.66 \\
        Baseline            & 0.0           & -0.0           & -0.0          & -0.0\\
        \toprule
    \end{tabular}
\end{table*}
\begin{table*}
    \caption{Relative quality difference (PSNR gain at the same bitrate; higher is better) and bitrate difference (bitrate gain for the same PSNR; lower is better) of the benchmarked methods on InStereo2k w.r.t. the baseline model. We also report the BD-Rate restricted to a low PSNR region ($32-36$dB) and high PSNR region ($38-40$dB).}
    \label{tab: instereo bd table}
    \centering
    \vspace{0.3cm}
    \begin{tabular}{lcccc}
        \toprule
        Method & BD-PSNR [dB]$\uparrow$ & BD-Rate [\%]$\downarrow$ & \tb{BD-Rate [\%]$\downarrow$}{low PSNR} &\tb{BD-Rate [\%]$\downarrow$}{high PSNR} \\
        \hline
        % These numbers are with respect to BPG vvvv
        ECSIC (proposed)    & \bf{0.77}     & \bf{-19.96}    & \bf{-37.04}   & \bf{-12.56} \\
        No context modules  & 0.63          & -18.20         & -27.67        & -9.61 \\
        Only decoder SCA    & 0.32          & -9.36          & -15.57        & -5.56 \\
        Only encoder SCA    & 0.0           & -0.70          & -2.53        & -0.81 \\
        Baseline            & 0.0           & -0.0           & -0.0        & -0.0\\
        \toprule
    \end{tabular}
\end{table*}
% \begin{figure*}[h!]
%     \centering
%     \hfill\includegraphics[width=.5\linewidth]{figures/cityscapes_ms_ssim.pdf}\hfill%
%     \includegraphics[width=.5\linewidth]{figures/instereo_ms_ssim.pdf}\hfill\null%
%     \caption{Rate-distortion curves for our method against other codecs for Cityscapes (left column) and InStereo2K (right column) measured by MS-SSIM.}
%     \label{fig: rd curves ms-ssim}
% \end{figure*}
\begin{figure*}
    \centering
    \setlength{\tabcolsep}{1pt}
        \hfill\includegraphics[width=.5\linewidth]{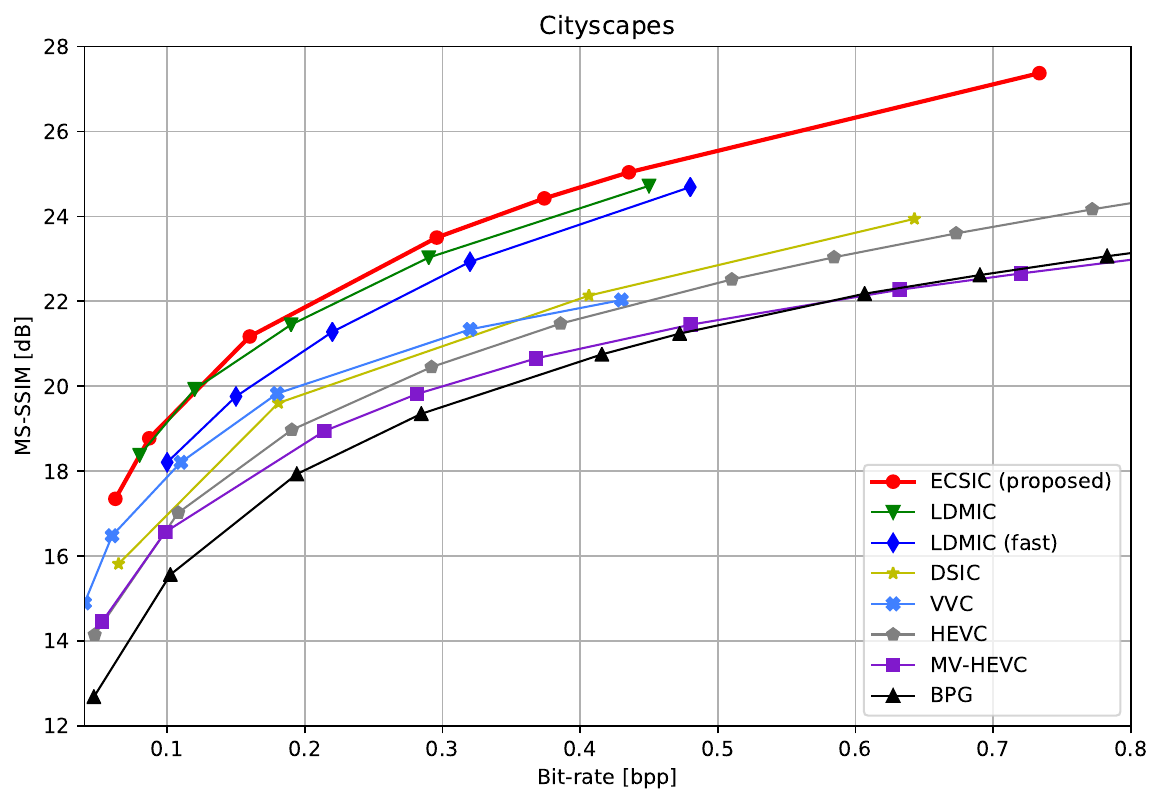}\hfill%
    \includegraphics[width=.5\linewidth]{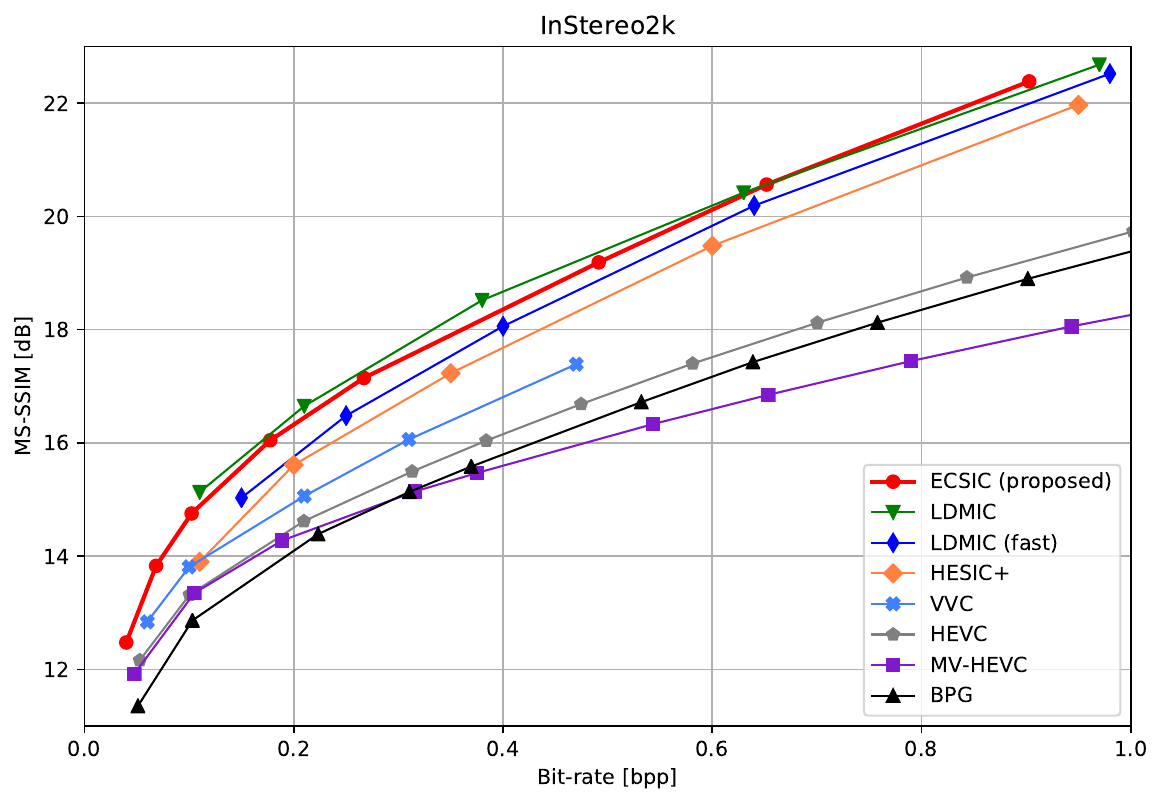}\hfill\null%
    % \begin{tabular}{c}
    %     \includegraphics[width=0.45\linewidth]{figures/cityscapes_ms_ssim.pdf}\\
    %     \includegraphics[width=0.45\linewidth]{figures/instereo_ms_ssim.pdf}
    % \end{tabular}
    \caption{Rate-distortion curves for our method against other codecs for Cityscapes (left column) and InStereo2K (right column) measured by MS-SSIM.}
    \label{fig: rd curves ms-ssim}
\end{figure*}

\subsection{MS-SSIM Rate-Distortion Curves}
\label{subsec: ms-ssim}
We provide MS-SSIM\cite{msssim} rate-distortion curves on Cityscapes and InStereo2k. For this, we initialize the model with an MSE pretrained model and fine-tune it with an MS-SSIM distortion loss. In the initial $50$k of the fine-tuning steps we train with a convex combination of MS-SSIM and MSE and linearly change the weight from MSE to MS-SSIM. After this initial warm-up phase we train with the loss
\begin{equation}
    \mathcal{L} = \mathcal{R} + \lambda \cdot (1 - \operatorname{MS-SSIM}(\bm{x}, \bm{\hat{x}}))
\end{equation}
where $\mathcal{R}$ denotes the rate loss term from eq. (6) from the main text, $\lambda$ denote the MS-SSIM rate/distortion trade-off parameter and $\bm{x}$ and $\bm{\hat{x}}$ denote the true and predicted stereo image pair respectively. The value for $\lambda$ was chosen such that the resulting model matches the bitrate of the MSE pretrained model. We train with a learning rate of $10^{-5}$ for ~600k steps epochs on Cityscapes and ~1M steps on InStereo2k. The result can be seen in Fig.~\ref{fig: rd curves ms-ssim} for Cityscapes on the left and for InStereo2k on the right. Our method outperforms all other methods on Cityscapes and is only slightly worse than the significantly slower (due to its autoregressive component) LDMIC model on InStereo2k.
% \begin{figure}
%     \centering
%     \includegraphics[width=\linewidth]{figures/cityscapes_ms_ssim.pdf}
%     \caption{MS-SSIM rate-distortion curves for our method against other codecs on Cityscapes.}
%     \label{fig: rd curves ms-ssim cityscapes}
% \end{figure}
% \begin{figure}
%     \centering
%     \includegraphics[width=\linewidth]{figures/instereo_ms_ssim.pdf}
%     \caption{MS-SSIM rate-distortion curves for our method against other codecs on InStereo2k.}
%     \label{fig: rd curves ms-ssim instereo}
% \end{figure}

\subsection{Submodules}
\label{subsec: submodules}
\begin{figure*}
	\centering
	\setlength{\tabcolsep}{1pt}
	\begin{tabular}{cl}
    	\includegraphics[width=.5\textwidth]{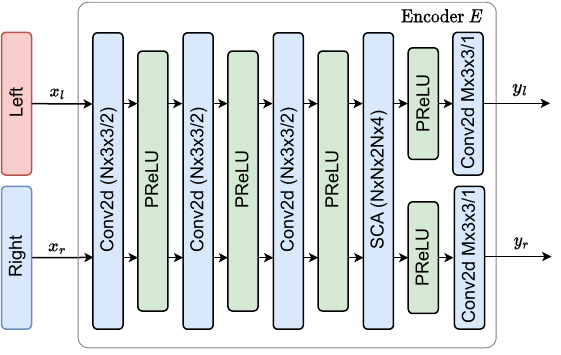} &
    	\includegraphics[width=.41\textwidth]{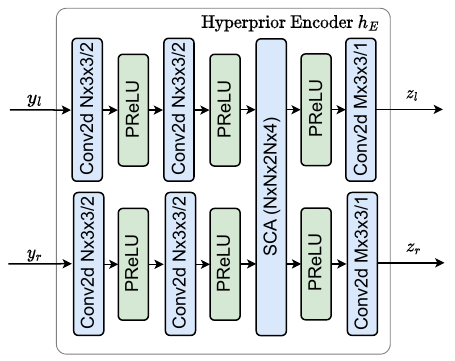}\\
    	\includegraphics[width=.5\textwidth]{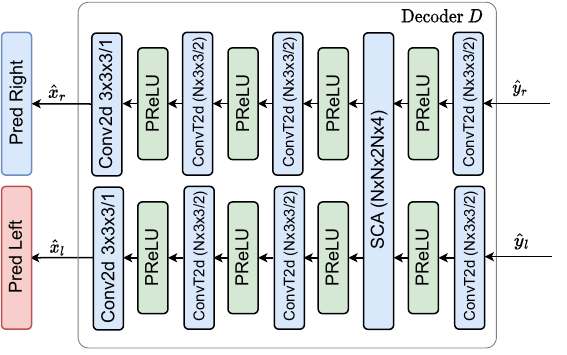} &
    	\includegraphics[width=.43\textwidth]{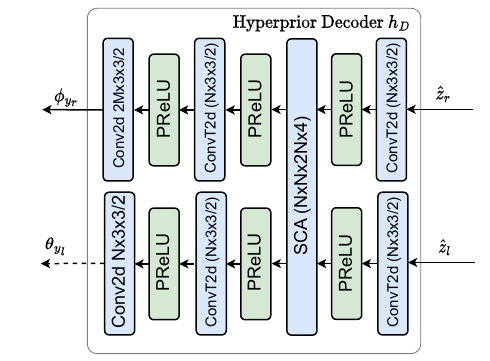}
	\end{tabular}
	\caption{The left columns shows encoder $E$ and decoder $D$. The right columns shows the encoder and decoder of the hyperior $h_E$ and $h_D$. We set $N = 192$ and $M = 48$ for all our experiments. Conv2d denotes 2d convolutional layers and ConvT2d 2d transposed convolutional layers. The initial three convolutional and PReLU layers in the encoder have shared weights between left and right stream.}
	\label{fig: submodules}
\end{figure*}

Fig.~\ref{fig: submodules} shows the layer structure of the encoder $E$ (top left), hyperprior encoder $h_E$ (top right), hyperprior decoder $h_D$ (bottom right) and decoder $D$ (bottom left). The encoder contains three downsampling steps and the hyperprior encoder contains two. The decoder and hyperprior decoder contain an equal amount of upsampling steps each. The initial three convolutional and PReLU layers in the encoder $E$ have shared weights between left and right input stream. In these modules the left and right streams are processed in parallel and are only connected in the SCA layers.

\subsection{Qualitative Results}
\label{subsec: qualitative results}
\begin{figure*}
	\centering
	\setlength{\tabcolsep}{1pt}
	\begin{tabular}{c}
    	\includegraphics[width=0.9\linewidth]{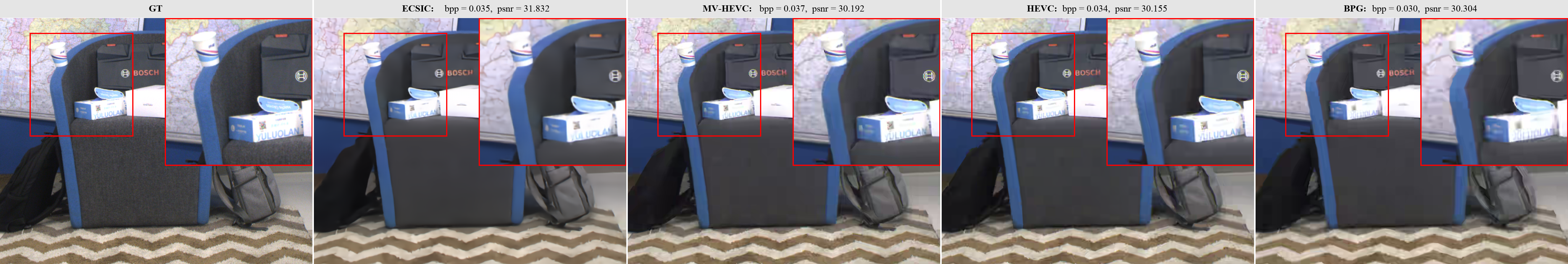}\\% max bpp_ratio
            \includegraphics[width=0.9\linewidth]{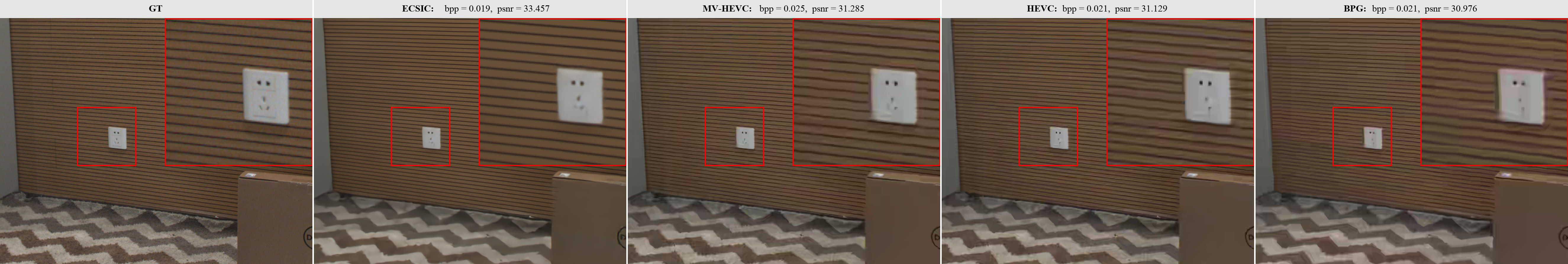}\\% min mse_ratio
    	\includegraphics[width=0.9\linewidth]{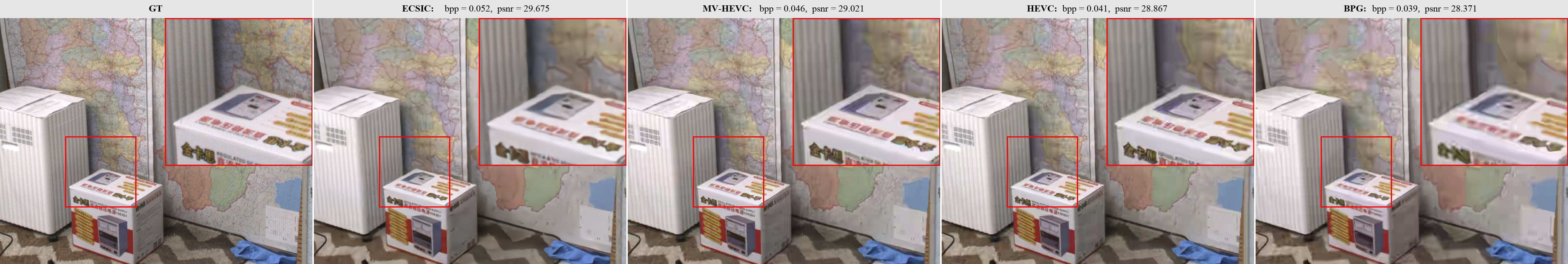}% max bpp
	\end{tabular}
	\caption{A qualitative comparison on images from the InStereo2K test set.}
	\label{fig: instereo comp}
\end{figure*}

\begin{figure*}
	\centering
	\setlength{\tabcolsep}{1pt}
	\begin{tabular}{cc}
    	\includegraphics[width=.5\textwidth]{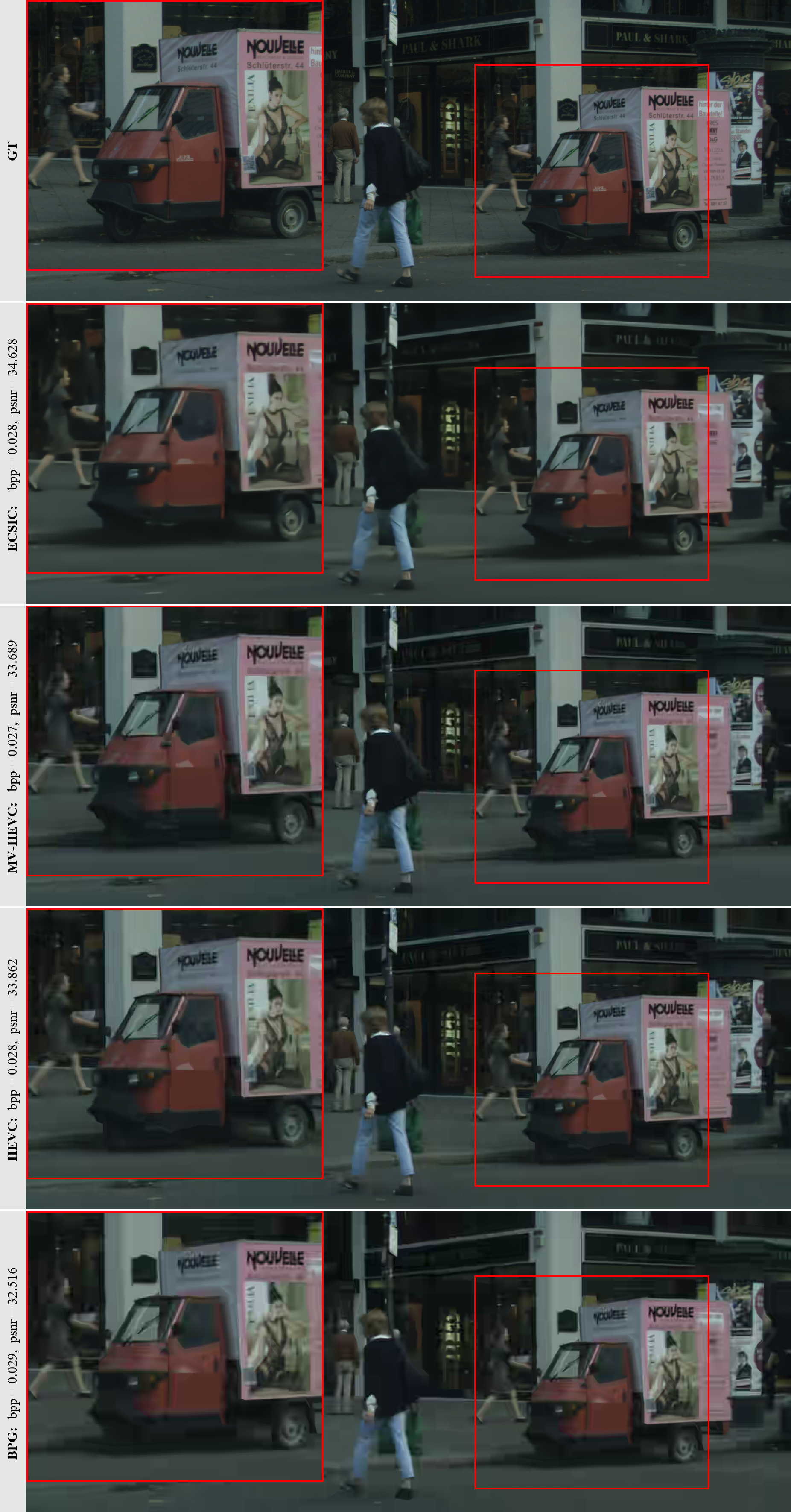} &%generic1
    	\includegraphics[width=.5\textwidth]{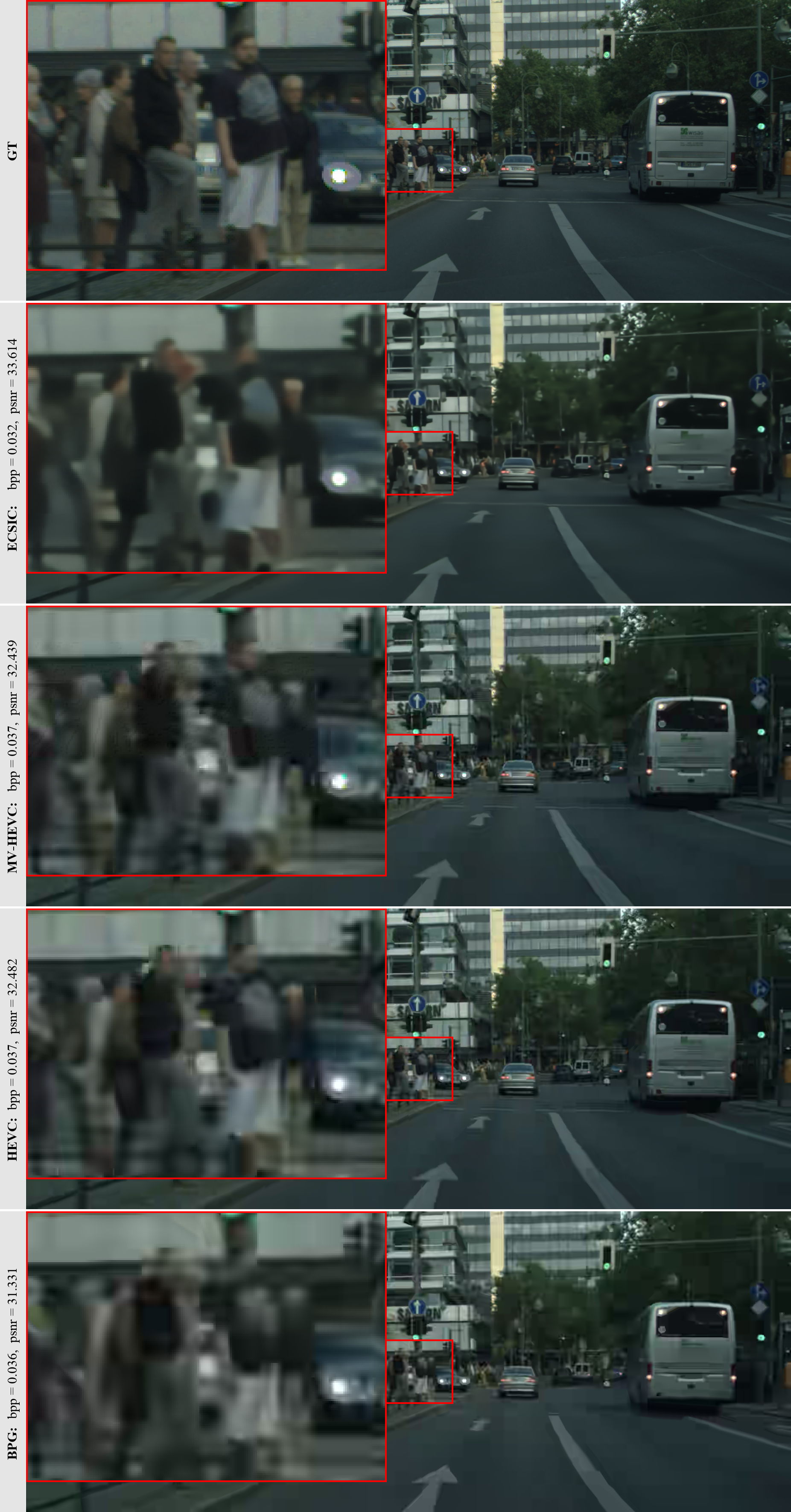}%generic2
	\end{tabular}
	\caption{A qualitative comparison on images from the Cityscapes test set.}
	\label{fig: cityscapes comp 1}
\end{figure*}

\begin{figure*}
	\centering
	\setlength{\tabcolsep}{1pt}
	\begin{tabular}{cc}
    	\includegraphics[width=.5\textwidth]{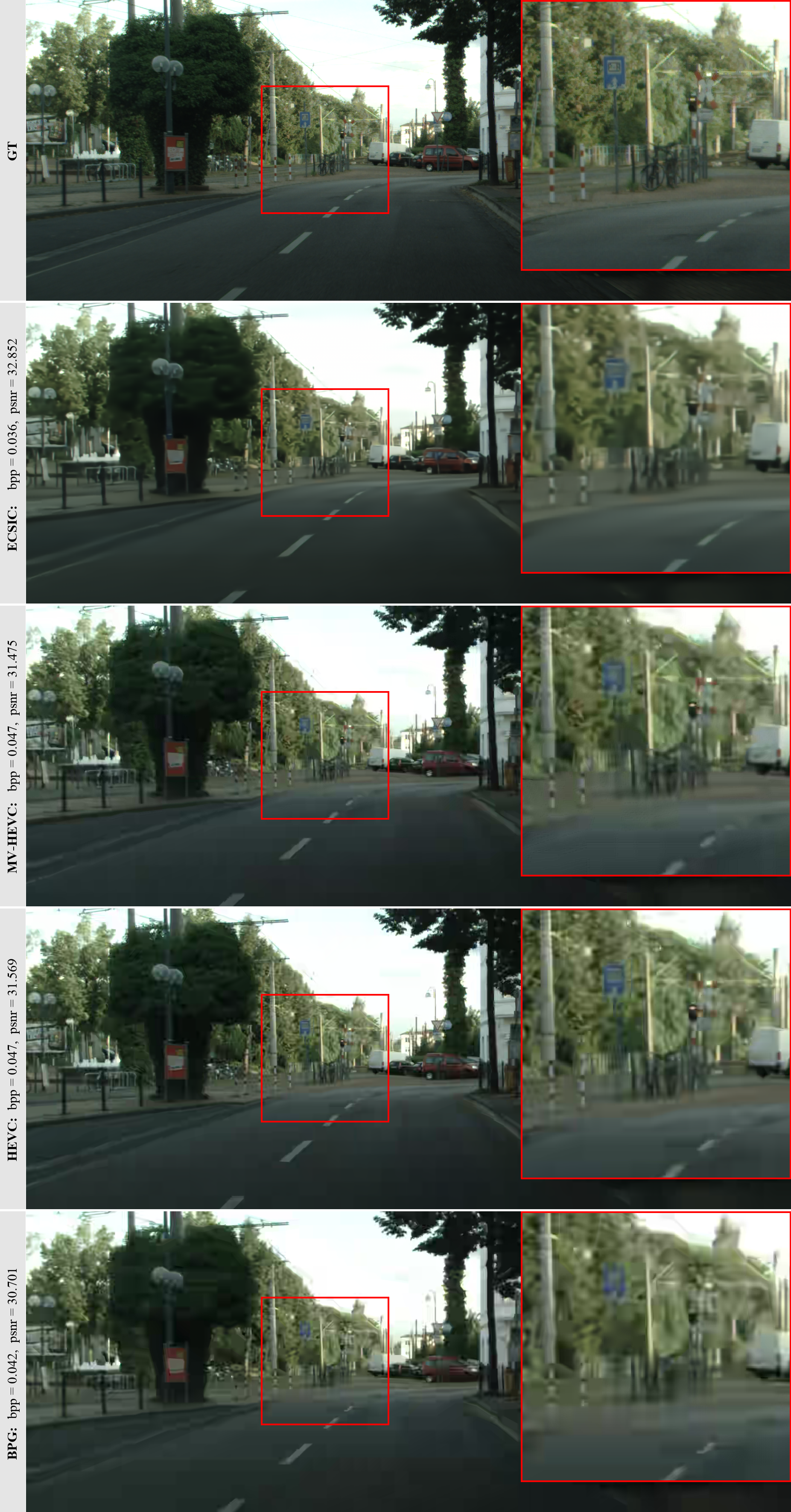} &%max mse_ratio
    	\includegraphics[width=.5\textwidth]{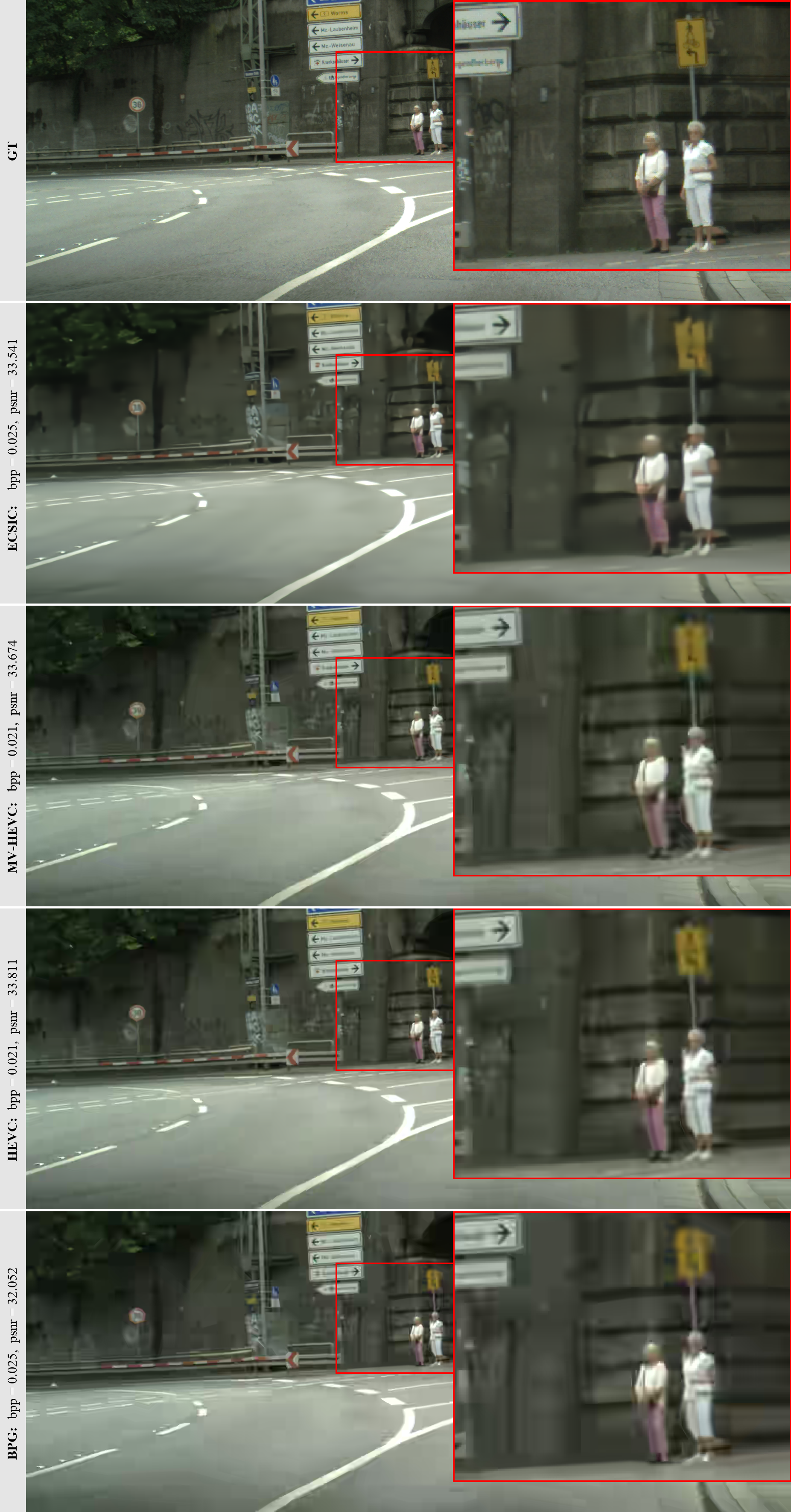}%min mse_ratio
	\end{tabular}
	\caption{A qualitative comparison on images from the Cityscapes test set.}
	\label{fig: cityscapes comp 2}
\end{figure*}

We provide additional qualitative results for examples from the InStereo2k dataset in Fig.~\ref{fig: instereo comp} and for Cityscapes in Fig.~\ref{fig: cityscapes comp 1} and Fig.~\ref{fig: cityscapes comp 2}. The example images always show the right image of the stereo image pair. Our method does not show noise patterns typical for traditional methods but instead results in a smoother appearance.


\end{document}